\documentclass[aps,pre,reprint,superscriptaddress,showpacs]{revtex4-1}

\usepackage{amsmath,amssymb}
\usepackage{bm}

\usepackage{color}

\usepackage{graphicx}
\graphicspath{{FIGS//}}

\begin{document}

\title{Kinetic Monte Carlo and Cellular Particle Dynamics Simulations of
  Multicellular Systems}

\author{Elijah Flenner} 
\affiliation{Department of Physics and Astronomy, University of Missouri,
  Columbia, MO 65211}

\author{Lorant Janosi}
\affiliation{Department of Physics and Astronomy, University of Missouri,
  Columbia, MO 65211}
        
\author{Bogdan Barz}
\affiliation{Department of Physics and Astronomy, University of Missouri,
  Columbia, MO 65211}

\author{Adrian Neagu} 
\affiliation{Department of Physics and Astronomy, University of Missouri,
  Columbia, MO 65211} 
\affiliation{Department of Biophysics and Medical Informatics, University of
  Medicine and Pharmacy Timisoara, 300041 Timisoara, Romania}

\author{Gabor Forgacs}
\affiliation{Department of Physics and Astronomy, University of Missouri,
  Columbia, MO 65211}
\affiliation{Department of Biological Sciences, University of Missouri,
  Columbia, MO 65211} 
\affiliation{Department of Biomedical Engineering, University of Missouri,
  Columbia, MO 65211} 

\author{Ioan Kosztin}
\email{kosztini@missouri.edu} 
\affiliation{Department of Physics and Astronomy, University of Missouri,
  Columbia, MO 65211}

\date{\today}

\begin{abstract}
  Computer modeling of multicellular systems has been a valuable tool for
  interpreting and guiding \textit{in vitro} experiments relevant to embryonic
  morphogenesis, tumor growth, angiogenesis and, lately, structure formation
  following the printing of cell aggregates as bioink particles.  Computer
  simulations based on Metropolis Monte Carlo (MMC) algorithms were successful
  in explaining and predicting the resulting stationary structures
  (corresponding to the lowest adhesion energy state). Here we present two
  alternatives to the MMC approach for modeling cellular motion and
  self-assembly: (1) a kinetic Monte Carlo (KMC), and (2) a cellular particle
  dynamics (CPD) method. Unlike MMC, both KMC and CPD methods are capable of
  simulating the dynamics of the cellular system in real time. In the KMC
  approach a transition rate is associated with possible rearrangements of the
  cellular system, and the corresponding time evolution is expressed in terms
  of these rates. In the CPD approach cells are modeled as interacting
  cellular particles (CPs) and the time evolution of the multicellular system
  is determined by integrating the equations of motion of all CPs. The KMC and
  CPD methods are tested and compared by simulating two experimentally well
  known phenomena: (1) fusion of two spherical aggregates of living cells, and
  (2) cell-sorting within an aggregate formed by two types of cells with
  different adhesivities
\end{abstract}

\pacs{%
87.17.Aa, 
87.17.Rt, 
87.85.G-, 
87.85.Lf  
}

\maketitle

\section{Introduction}
\label{sec:intro}
Understanding how living cells form tissues and organs is a fundamental
problem of developmental biology \cite{ingber07-2541,forgacs05}, and is also
important for the rapidly expanding field of tissue engineering that aims at
building functional tissue substitutes \emph{in vitro}
\cite{langer93-920}. Tissue engineered structures may be used for drug testing
and to restore or replace damaged tissues and organs
\cite{griffith02-1009}. An emerging tissue engineering technique is
bioprinting
\cite{mironov03-157,jakab04-2864,boland06-910,jakab08-413,mironov09-2164,norotte09-5910,jakab10-1} via
the automated layer-by-layer deposition of multicellular aggregates (the
bioink). Subsequent postprinting fusion of the contiguous aggregates gives
rise to the desired tissue construct.  Predicting the result of post-printing
tissue formation is a task for theoretical modeling.

A guiding principle for most models of cell rearrangement in cell aggregates
is the \emph{differential adhesion hypothesis} (DAH) proposed by Steinberg
\cite{steinberg63-401,steinberg70-395}.  DAH states that structure formation
in multicellular systems occurs due to (i) differences in cell-to-cell
adhesion of different types of cells and (ii) cell motility. Cells seek
positions with the largest number of strong bonds. For example, in a random
mixture of two cell types of different cohesivities the
more cohesive cell population sorts out and occupies the central region surrounded by
the less cohesive population.

By incorporating DAH, Metropolis Monte Carlo (MMC) simulations correctly 
predict the formation of multicellular structures of minimum energy of
adhesion, and identify long-lived, metastable configurations
\cite{neagu05-178104}.  However, MMC  cannot predict the actual
time evolution of multicellular systems.

Insight into time evolution emerged from experiments designed to verify DAH,
which revealed that embryonic tissues behave analogously to highly viscous
liquids.  The concept of tissue liquidity motivated a quantitative description
of embryonic tissues in terms of parameters of continuum hydrodynamics, such
as surface tension and viscosity. 
For example, the (apparent) surface tension $\gamma $ was measured for several tissue types using a parallel  plate compression apparatus \cite{foty94-2298} and the values were found to be consistent with the sorting behavior of these tissues \cite{foty96-1611}. 

Here, by using the concept of tissue liquidity and DAH, we formulate two alternatives to the MMC approach for modeling cellular motion and self-assembly: (1) a \emph{kinetic Monte Carlo} (KMC), and (2) a \emph{cellular particle dynamics} (CPD) method. {\color{black} Both these methods are capable of describing and predicting the real time evolution of the shape of multicellular systems and tissue constructs in certain morphogenetic processes and post-bioprinting structure formation.}
{\color{black} In the KMC method the configuration of the multicellular system is propagated in time through a standard rejection-free kinetic Monte Carlo algorithm. This approach should provide a more accurate description of the time evolution of a multicellular system than other grid based methods, such as, the MMC model \cite{jakab04-2864,neagu05-178104} or the widely used cellular Potts model (CPM).  The latter uses a modified Metropolis MC algorithm to update the configuration of the simulated system and postulates that time is proportional to the number of MC steps, which in general is not the case \cite{flenner08-461}. }
{\color{black} In the CPD method \cite{flenner08-461} individual cells are modeled as an ensemble of cellular particles (CPs) that interact via short range contact interactions, characterized by an attractive (adhesive interaction) and a repulsive (excluded volume interaction) component. CPs in a cell are held together by an additional confining potential that mimics the role of the cell membrane. The time evolution of the spatial conformation of the multicellular system is determined directly by recording the trajectories of all CPs by integrating their equations of motion. What sets apart CPD from the other similar off-grid particle methods, such as Newman's \textit{subcellular element method} (SEM) \cite{newman05-611,sandersius08-15002,sandersius11-45007,sandersius11-45008}, is the employed force field (especially the confining potential) and its parametrization that makes the system behave as a complex viscous liquid. In particular, the CPD model parameters are determined such that the shape of two fusing spherical aggregates in the CPD simulation match as closely as possible the one observed experimentally, i.e., two attached spherical caps (see Fig.~\ref{caps}) \cite{jakab08-2438}. }
To test and compare the KMC and CPD methods, we apply them to simulate the fusion of two spherical aggregates and the evolution of cell sorting within an aggregate, two morphogenetic processes driving postprinting structure formation.
{\color{black} For the theoretical description of the fusion of two identical spherical aggregates we use a simple continuum model introduced by Frenkel \cite{frenkel45-385} and further developed by others working in the field of rheology \cite{pokluda97-3253,bellehumeur98-270}.   }
It is this theoretical continuum model that provides the link between the time scales of simulations and the time scales of experiments. Once this link is established, the KMC and CPD simulations are used to quantitatively predict the time evolution of complex postprinted structures whose description using a continuum hydrodynamics approach is impractical.

The remainder of the paper is organized as follows. Section~\ref{sec:theory} describes the KMC (Sec.~\ref{sec:kmc}) and the CPD (Sec.~\ref{sec:cpd}) methods, as well as the theoretical aspects of the continuum approach of aggregate fusion (Sec.~\ref{sec:cont}).  Section~\ref{sec:results} contains the results and discussion of our KMC and CPD simulations, i.e., fusion of identical spherical multicellular aggregates (Sec.~\ref{sec:fusion}) and cell sorting (Sec.~\ref{sec:sorting}). Conclusions are presented in Sec.~\ref{sec:conclusions}.

\section{Computer and Theoretical Modeling}
\label{sec:theory}

\subsection{Kinetic Monte Carlo for Multicellular Systems}
\label{sec:kmc}
The Kinetic Monte Carlo method (KMC) was proposed as an alternative to the MMC
method for simulating the evolution of Ising models \cite{bortz75-10}.  When a
system approaches equilibrium, or is in a metastable state, the Metropolis
algorithm rejects most trial moves because the acceptance probability is
small. A main feature of the KMC algorithm is that it is ``rejection-free''.
In each step, one calculates the transition rates for all possible changes
compatible with the current configuration, and then chooses a new
configuration with a probability proportional to the rate of the corresponding
transition.
 
We designed and implemented a KMC algorithm to simulate the time
evolution of a lattice model of multicellular systems.  Aggregates of
cells in cell culture medium are represented on a 3D hexagonal
close-packed lattice by associating each site to either a cell or to a similar
sized volume element of medium. Thus, the lattice spacing is equal to one cell
diameter.  We assume that each cell interacts with its 12 nearest neighbors (1st and 2nd neighbors considered to be nearest neighbors) located at a distance of one lattice spacing from the given cell. lnteractions are expressed in terms of works of cohesion and adhesion
\cite{israelachvili97,foty04-397}, defined as the work needed to break up the
contact between two neighbors of respectively similar or differing types of cells. For example, in case of a multicellular aggregate composed of a single cell type, the work needed to extract a cell from the aggregate (i.e. model tissue)  is the work of cohesion, $\epsilon_{cc}$, multiplied by the number of the cell's nearest neighbor. The
interaction between cells and the cell culture medium is set to zero.
The movement of cells is described by assigning rates to swapping cells with adjacent cells of different type and/or with medium elements. These elementary moves occur with rates
given by
\begin{equation}
k = w_0 e^{-E_{\rm b}/E_T}, 
\label{rates}
\end{equation}
where the factor $w_0$ is the frequency of attempts to cross the energy
barrier of height $ E_{\rm b}$, and $E_T$ is the energy of biological
fluctuations \cite{beysens00-9467}, the analog of the energy of thermal fluctuations, $k_B
T$  ($k_B$ is Boltzmann's constant  and  $T$ is the absolute temperature). It has been argued that $E_T$ is a characteristic measure of cell
motility: the higher is $E_T$ in comparison to the energies of
cohesion/adhesion, the higher is the motility of the cell 
\cite{beysens00-9467}.

Due to the complexity of the cytoskeletal machinery responsible for cell
movement, there is no unique way to assign a barrier height to the swapping of
two cells.  Any reasonable choice, however, needs to be consistent with the
following set of experimental observations on cell movement in 3D:

\begin{itemize}
\item[(1)] Relocation of cells in embryonic tissues and in some engineered tissues (such as cell aggregates) occurs according to DAH  \cite{steinberg70-395,Steinberg82}: cells take advantage of their motility to establish the maximum number of strong bonds with their neighbors. 

\item[(2)] Anchorage-dependent cells do not spontaneously dissociate from the cell aggregate they are part of  \cite{steinberg70-395}. 

\item[(3)] The speed of cell movement in 3D matrices has a particular dependence on the strength of cell-matrix adhesion: cell movement is fastest at an optimal strength of binding. Too weak or too strong binding hampers cell movement \cite{palecek97-537, zaman05-1389}. 

\end{itemize} 

Consider a binary particle model for a multicellular system formed by two cell types, $t=1,2$ (for a multicellular aggregate composed of one cell type, $t=1$, surrounded by tissue culture medium, $t=2$ represents the medium particle). The configurational energy (or total interaction energy), $E$ is expressed as \cite{jakab04-2864}
\begin{equation}
\label{eq:totalenergy}
E=\gamma_{12} N_{12}+{\it const}, 
\end{equation}
where
$\gamma_{12} = (\epsilon_{11} + \epsilon_{22})/2 - \epsilon_{12}$, with $\epsilon_{11}$ and $\epsilon_{22}$ being the energies of cohesion respectively for cell type 1 and 2, and $\epsilon_{12}$ is the energy of adhesion. $N_{12}={\sum_{i=1}^{N_{1}^I}{n_{i2}}}={\sum_{i=1}^{N_{2}^{II}}{n_{i1}}}$ is total number of nearest neighbor pairs of different cell types cells, $n_{i2}$ ($n_{i1}$)  
the number of nearest neighbors of cell $i$ of type 1(2), which are of type 2(1) and $N_{1}^{I}$ ($N_{2}^{II}$) the total number of cells of type 1(2), which have at least one (nearest) neighbor of type 2(1).   
(As the {\it const}  is irrelevant for the evolution of the system, we set it to zero \cite{jakab04-2864}.)

Consider two nearest neighbor cells, $i$ and $j$ of different types (without
loss of generality we can set $i=1$ and $j=2$). The system evolves in time
towards configurations of decreasing energy $E$, i.e. for $\gamma_{12}>0$
($\gamma_{12}<0$) $N_{12}$ decreases (increases). For $\gamma_{12}>0$ and
$\gamma_{12}<0$ cells respectively phase separate (cell sorting) and mix (cell
mixing).  Elementary KMC moves consist of swapping two neighbors of different
types (swapping cells of same type does not change the energy). The
contribution of two such cells, $i$ and $j$ to $E$ is
\begin{equation}
E_{ij}=\frac{1}{2}(n_{i2}+n_{j1}) ,
\label{energy_pair}
\end{equation}
and $E={\sum_{i=1}^{N_{1}^I}} {\sum_{j=1}^{N_{2}^{II}}}E_{ij}$. Furthermore, the larger is $E_{ij}$ the more likely is the KMC move to swap cells $i$ and $j$. Thus it is reasonable to define the energy barrier  $E_{\rm b}^{ij}$ in Eq.~(\ref{rates}), for a transition involving the swapping of two cells $i$ and $j$, as 
\begin{equation}
0{\le}E_{\rm b}^{ij}= E_{ij}^{max}-E_{ij} ,
\label{barrier_height}
\end{equation}
where $E_{ij}^{max}$ is the maximum possible value of $E_{ij}$.
For $\gamma_{12}>0$, $E_{ij}^{max}$ is obtained when the number of neighbors
of differing type surrounding cells $i$ and $j$ is maximal.

Now we can formulate the steps of our KMC algorithm for simulating the time
evolution of multicellular systems: (S1) Set $t=0$; (S2) Find all interfacial
cells (i.e., cells in contact with cell culture medium or with cells of
different type) and compute the rates $k_{m}$, $1\le m \le M$, corresponding
to all possible $M$ transitions involving these cells; (S3) Calculate the
cumulative rates: $K_m = \sum_{n=1}^{m} k_n$, $1\le m \le M$; (S4) Generate a
uniform random number $u$ between $0$ and $1$ and carry out event ``$m$'' for
which $K_{m-1} < u K_M \le K_m$; (S5) Generate another uniform random number
$u'$ between $0$ and $1$, and increment the time variable (i.e., $t\rightarrow
t+\Delta{t}$) by the non-uniform time step
\begin{equation}
  \label{eq:dt}
\Delta t = -K_M^{-1} \log(u') ;
\end{equation}
(S6) Update all rates $k_n$ that may have changed due to the previous
transition ``$m$''; (S7) Return to step S2 and repeat the process until the
time variable reaches the desired target
value. 

\subsection{Cellular Particle Dynamics Method for Multicellular Systems}
\label{sec:cpd}

{\color{black} The \emph{cellular particle dynamics} (CPD) method is an off-lattice, particle-based computer simulation method that can describe and predict the time evolution of 3D multicellular systems during shapechanging biomechanical transformations \cite{flenner08-461}. 
Within the CPD formalism cells, regarded as continuous objects with self-adaptive shape, are coarse-grained into a finite number, $N_{CP}$, of equal volume elements. Each volume element is represented by a point-like cellular particle (CP) situated at its center of mass. CPs interact via short-range contact interactions, characterized by an attractive (adhesive interaction) and a repulsive (excluded volume interaction) component. In addition, CPs within a given cell are subject to a confining potential that assures the integrity of the cell. The time evolution of the spatial conformation of the multicellular system is determined directly by calculating the trajectories of all CPs (and, therefore, cells) through integration of their overdamped Langevine equations of motion. This minimalist model, when properly parametrized, has the features of a complex viscous liquid and it is suitable for describing the time evolution of multicellular aggregates and soft-tissue constructs.

For the $n^{\text{th}}$ CP in cell $\alpha$, the equation of motion is }
\begin{equation}
 \label{eq:cpd-eq-motion}
 \mu \dot{\bm{r}}_{\alpha_n}(t) = -\nabla_{\alpha_n} U + \bm{f}_{\alpha_n}(t),
\end{equation}
where $\bm{r}_{\alpha_n}(t)$ is the position vector, $U$ is the potential
energy function describing the interaction of the CPs , $\mu$ is the friction
coefficient, $\bm{f}_{\alpha_n}(t)$ is a random force, and the dot denotes
time derivative.  The random force is modeled as a Gaussian white noise with zero
mean and variance $\langle f_i(t) f_j(0) \rangle = 2 D \mu^2 \delta(t)
\delta_{ij}$, where $D$ is the sort-time self diffusion coefficient of the
CPs. The CPD parameters $D$ and $\mu$ are related to the previously introduced
biological fluctuation energy $E_T$ by the Einstein relation $D\mu=E_T$.
The CP potential energy $U$ has an intra-cellular and an inter-cellular
component corresponding to CPs belonging respectively to the same cell and to
different cells, i.e.,
\begin{eqnarray}
\label{eq:cpd-U}
U  & = & \frac{1}{2} \sum_{\alpha} \sum_{n=1 \atop m \ne n} U^{intra}(|\bm{r}_{\alpha_n}-\bm{r}_{\alpha_m}|) \nonumber \\
 & & + \frac{1}{2} \sum_{\alpha \atop \beta \ne \alpha} \sum_{n,m} U^{inter}(|\bm{r}_{\alpha_n} - \bm{r}_{\beta_m}|),
\end{eqnarray}
where $\alpha_n$ ($\beta_m$) labels the cellular particle $n$ ($m$) in cell $\alpha$
($\beta$).
{\color{black}We model the short-range contact inter- and intra-cellular interactions between CPs through 
\begin{subequations}
\label{eq:cpd-pot}
\begin{equation}
\label{eq:Uinter}
U^{inter}(r) = V_{LJ}\left(r;\epsilon^{inter},\sigma^{inter}\right) ,\\
\end{equation}
\begin{equation}
\label{eq:Uintra}
U^{intra}(r) = V_{LJ}\left(r;\epsilon^{intra},\sigma^{intra}\right) +  \frac{k}{2} \left( r - \xi \right)^2 \Theta(r-\xi) , \\
\end{equation}
where
\begin{equation}
\label{eq:LJ}
  V_{LJ}(r;\epsilon,\sigma) = 4\epsilon\left[\left(\frac{\sigma}{r}\right)^{12} - \left(\frac{\sigma}{r}\right)^{6} \right]
\end{equation}
\end{subequations}
is the standard Lennard-Jones potential, and $\Theta(r)$ is the Heaviside step function. Note that instead of $V_{LJ}$ one could use any other potential that has a repusive core and a short range attractive part. For example, in SEM \cite{newman05-611} a Morse potential is used to describe the interaction between two subcellular elements.}
However, the important addition in CPD is the second quadratic term in $U^{intra}$ that, for $r > \xi$,  represents an \emph{elastic confining potential} used to maintain the integrity of the cell. This term, characterized by the elastic constant $k$, guarantees that the CPs within a cell remain confined inside the boundary of the cell.
The time evolution of the multicellular system within the CPD approach is determined by numerically integrating the equations of motion (\ref{eq:cpd-eq-motion}) for all CPs. We have accomplished this by implementing the intra- and inter-cellular interaction forces, Eqs.~(\ref{eq:cpd-U})-(\ref{eq:cpd-pot}), and a Langevin dynamics integrator in the freely available massively parallel molecular dynamics package LAMMPS \cite{plimpton95-1}.

{\color{black}
  \subsubsection{CPD units and parameters}
  \label{sec:cpd-units}

For a multicellular system of a single cell type there are nine CPD model parameters that need to be determined: $N_{CP}$ (the number of CPs per cell), $D$, $\mu$, $\sigma^{intra}$, $\varepsilon^{intra}$, $k$, $\xi$, $\sigma^{inter}$ and $\varepsilon^{inter}$. 
The choice of $N_{CP}$ is determined by the degree of detail we want to describe individual cells. Since  we are interested in the time evolution of the shape of an aggregate formed by a large number of cells and not in the detailed description of the surface dynamics of individual cells, in the present work we make the reasonable choice of $N_{CP}=10$. 
Because $\sigma$ in \eqref{eq:LJ} determines the size of the interacting CPs, we can set $\sigma\equiv\sigma^{intra}=\sigma^{inter}$. 
The length $\xi$ in \eqref{eq:Uintra} represents the size (diameter) of a cell, which comprises $N_{CP}$ tightly packed CPs of size $\sigma$. Thus, one can estimate $\xi\approx\sigma N_{CP}^{1/3}$.

Next, we define convenient CPD (or computer) length, time and energy units according to 
\begin{equation}
  \label{eq:cdp-units}
\ell_0 = \sigma ,\quad t_0=\frac{\sigma^2}{D} ,\quad E_0=E_T=\mu D .
\end{equation}
In these units all CPD parameters all pure numbers, and in particular $\sigma=D=\mu=1$. We set the confining potential parameters as $\xi=2.5$ $(\sim 10^{1/3})$ and $k=5$. A larger (smaller) value for $k$ makes the cell more rigid (soft) when subjected to deformations. The chosen value, for which $k\sigma^2/2=2.5 E_T$, is suitable when cells in the aggregate are exposed only to adhesion and surface tension forces. By choosing $\epsilon^{intra}=1$ (i.e., same as the biological fluctuation energy $E_T$) the dynamics of the CPs inside a cell will have sufficient randomness to produce cell surface fluctuations that play an important role in cell motility \cite{gordon72-43}.  

Thus, out of the nine CPD parameters we are left with only one, $\epsilon^{inter}$, that needs to be determined such that the time evolution of the shape of the multicellular system follows as closely as possible the corresponding experimental one. For this purpose we focus on the fusion of two identical spherical aggregates, as described in the next section. Based on extensive CPD simulations we have found that the best agreement with experiment is obtained for $1<\epsilon^{inter}<2$, when the system behaves as a viscous liquid. By increasing $\epsilon^{inter}$ above $2$ the fusing cellular aggregates show sign of solidification and their behavior deviate significantly from experiment.
The results reported in this paper are for $\epsilon^{inter}=1$. However, these are similar to the ones obtained for any  $\epsilon^{inter}< 2$.

Furthermore, in all our CPD simulations we have used an integration time step
$\Delta{t}=10^{-4}$, and used a cutoff radius $R_c=2.5$ for $U^{inter}(r)$.
}

\subsection{Continuum Description of the Fusion of Two Spherical Cell Aggregates}
\label{sec:cont}

\begin{figure}
\includegraphics[width=3.2in]{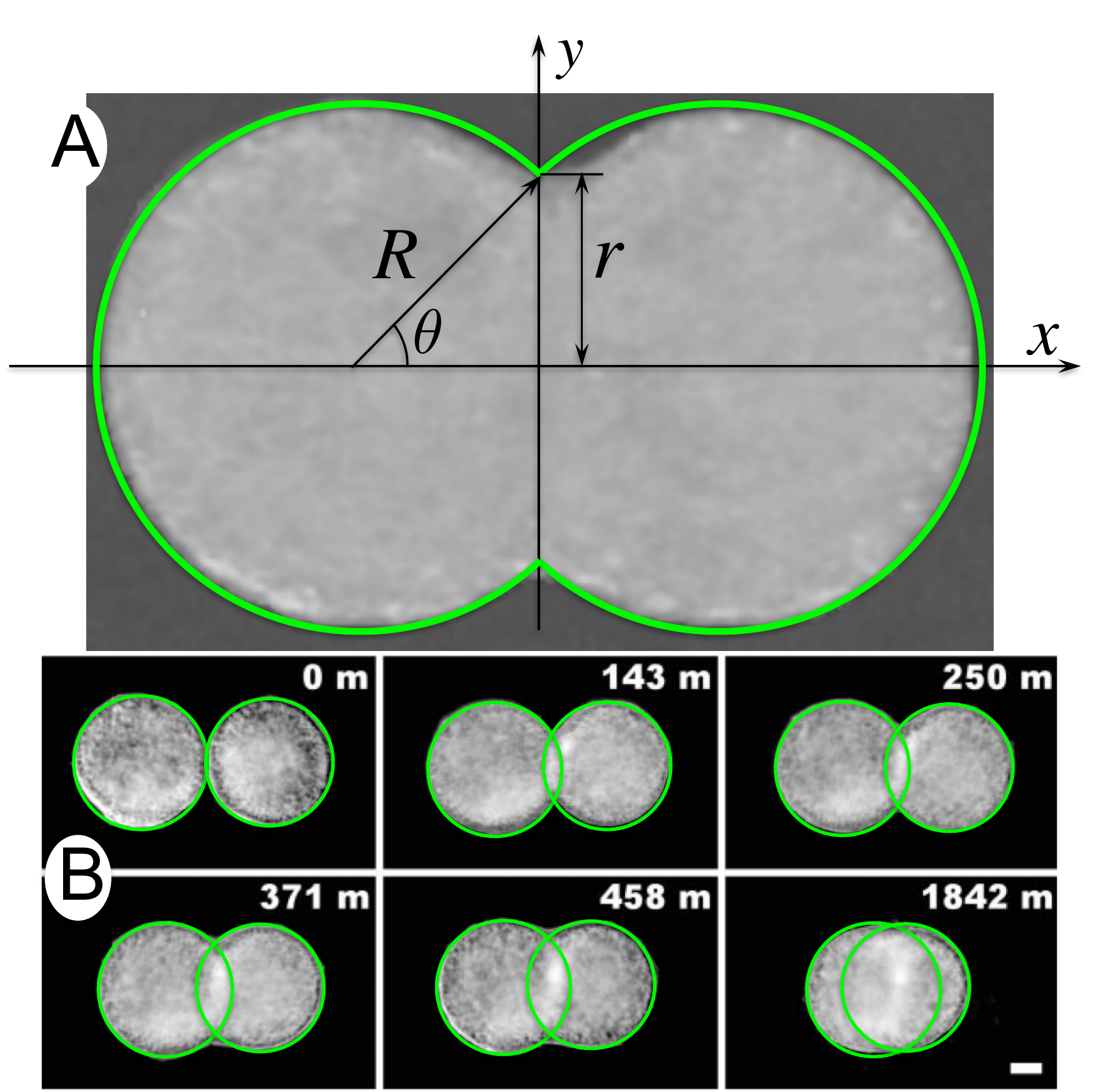}
\caption{\label{caps} (color online). (A) The shape of two fusing
  spherical cell aggregates can be quantified by the angle $\theta(t)$
  and the radius $R(t)$. (B) Throughout the fusion of two indentical
  spherical cushion tissue aggregates \cite{jakab08-2438} the system
  has the shape of two connected spherical caps. The numbers indicate
  the time (in minutes) elapsed from the start of the fusion.}
\end{figure}

{\color{black} The fusion of two contiguous cell aggregates is driven by surface tension, $\gamma$, and resisted by viscosity, $\eta$.  
It is an experimental fact that during the fusion of identical spherical soft tissue aggregates the shape of the system is that of two touching spherical caps (see Fig.~\ref{caps}) \cite{jakab08-2438}. This observation suggests that soft tissues behave like complex viscous liquids whose description requires an a priori unknown hydrodynamic constitutive model. However, the simplicity of the geometry allows us to describe analytically the dynamics of the considered fusion process by employing conservation laws as proposed by Frenkel \cite{frenkel45-385} and Eshelby \cite{eshelby49-806} for the coalescence (sintering) of highly viscous molten drops. }

The fusing aggregates are modeled as two spherical caps of radius $R(\theta)$ with circular contact (``\textit{neck}'') region of radius $r(\theta)=R(\theta)\sin\theta$ (see Fig.~\ref{caps}A). Volume conservation requires
\begin{equation}
  \label{eq:Rt} R(\theta) = 2^{2/3}(1+\cos\theta)^{-2/3} (2 -\cos\theta)^{-1/3} R_0 \;,
\end{equation}
with $R_0=R(0)$. Thus, the time evolution of the fusion process is parametrized by a single angle $\theta=\theta(t)$, defined in Fig.~\ref{caps}A, that changes from $\theta(0)=0$ to $\theta(\infty)=\pi/2$.
The rate of the decrease in surface energy is $\dot{W}_s = \gamma dS/dt$, where the free surface area $S=S(\theta) = 4 \pi R^2(\theta) (1+\cos\theta)$.
The equation of motion for $\theta(t)$ can be derived by equating $\dot{W}_s$ with the rate of the energy dissipated by the viscous flow $\dot{W}_{\eta} \approx -4 \pi R_0^3 \eta \alpha^2$ \cite{frenkel45-385,bellehumeur98-270}. Assuming biaxial stretching flow,

 \begin{equation}
  \label{eq:alpha}
 \alpha =\frac{\partial v_x}{\partial x} \approx -\frac{1}{R(\theta)}
 \frac{d}{dt} [R(\theta) \cos\theta] .
 \end{equation} 
Inserting Eq.~(\ref{eq:alpha}) into the energy balance equation
$\dot{W}_s=\dot{W}_{\eta}$ leads to \cite{bellehumeur98-270}

 \begin{equation}
\frac{d\theta}{dt}=\frac{1}{\tau} \frac{\sin\theta \cos\theta
  (2-\cos\theta)^{1/3}}{2^{5/3}(1-\cos\theta)(1+\cos\theta)^{1/3}} =
\frac{1}{2\tau} \frac{R_0\cot\theta}{R(\theta)} ,
 \label{r0-diff-eq}
 \end{equation}
where the characteristic fusion time
\begin{equation}
  \label{eq:tau}
  \tau = \eta R_0/\gamma \;.
\end{equation}

Equation (\ref{r0-diff-eq}) can be solved numerically for $\theta=\theta(t)$. 
However, one can derive a simple and accurate analytical
approximation for $\theta(t)$ by setting $R(\theta)\approx R_0$ in
Eq.~(\ref{r0-diff-eq}). Indeed, throughout the fusion process $1 \leq
R(\theta)/R_0 \leq 2^{1/3}\approx 1.26$ holds. With this approximation,
Eq.~\eqref{r0-diff-eq} can be easily integrated with the result
\begin{equation}
  \label{eq:costheta}
  \cos\theta = \exp(-t/2\tau) \;.
\end{equation}
Note that according to Eqs.~(\ref{r0-diff-eq}) and (\ref{eq:costheta}) the
dynamics of the fusion process, described by $\theta(t)$ as a function of
$t/\tau$, is independent of the size (i.e., $R_0$) of the fusing
spheres. $R_0$ appears only in the characteristic fusion time $\tau$,
Eq.~(\ref{eq:tau}).

Finally, using Eq.~(\ref{eq:costheta}), the square of the radius of the
circular neck region of the fusing spherical caps can be expressed as
\begin{subequations}
\label{exp_fit_formula}
\begin{equation}
\left( \frac{r}{R_0} \right)^2 \approx A(t) [1-\exp(-t/\tau)]\;,
\label{exp_fit_formula-1}
\end{equation}
with
\begin{equation}
  \label{exp_fit_formula-2}
  A(t) = 2^{4/3}\left(1+e^{-t/2\tau}\right)^{-4/3} \left(2-e^{-t/2\tau}\right)^{-2/3}\;.
\end{equation}
\end{subequations}

{\color{black} For short times, $t\ll\tau$, Eqs.~(\ref{exp_fit_formula}) yield the familiar linear-in-time expression, $(r/R_0)^2\approx t/\tau$, obtained by Frenkel \cite{frenkel45-385} and Eshelby \cite{eshelby49-806}. While this formula has been applied previously to estimate the capillary velocity $v_c=\gamma/\eta=R_0/\tau$ of soft tissues \cite{gordon72-43,jakab08-2438}, we are not aware of any previous study that followed the time evolution of the shape and of $[r(t)/R_0]^2$ throughout the fusion process of two spherical tissue aggregates. First, we have determined the dimensionless CPD parameters (i.e., expressed in CPD units; see Sec.~\ref{sec:cpd-units}) such that the shape of the fusing aggregates during CPD simulation resemble as close as possible to spherical caps. Second, we have determined the characteristic fusion time $\tau$ by fitting the data for $[r(t)/R_0]^2$, obtained respectively from experiment, CPD and KMC simulations, to Eqs.~(\ref{exp_fit_formula}). Finally, the CPD time unit can be calculated as  $t_0=\tau_{exp}/\tau_{CPD}$. Once $t_0$ is known, one can predict through CPD simulation the time evolution of an arbitrary 3D tissue construct built from the same type of cells for which the time calibration was performed through the above method (i.e., fusion of spherical aggregates). Clearly a similar calibration strategy can be used for the KMC method.
}

\section{Results and Discussion}
\label{sec:results}

To test and compare the KMC and CPD methods described in Sec.~\ref{sec:theory}, we have applied them to simulate two important morphogenetic processes: (A) tissue fusion (the fusion of two identical spherical multicellular aggregates), and (B) cell-sorting (within a spherical multicellular aggregate formed by two types of cells with different adhesivities).

\subsection{Fusion of Two Spherical Cell Aggregates}
\label{sec:fusion}
As descirbed in Sec.~\ref{sec:cont}, the fusion of two identical spherical
aggregates can quantitatively be characterized by the time dependence of the
radius, $r(t)$, of their circular contact region. According to
Eqs.~(\ref{exp_fit_formula}), $r(t)$ obtained from experiment and from KMC and
CPD simulations, can be used to determine the characteristic fusion time
$\tau$, Eq.~(\ref{eq:tau}). Thus, for a given cell type, by comparing the
experimental $\tau$ with that obtained from computer simulations one can
calibrate the time scale of the corresponding computer model. Once such a
calibration is done, one can make quantitative \textit{in silico} predictions
of the time evolution of various multicellular processes that involve the same
cell type \cite{flenner08-461}.

In this section we present KMC and CPD simulation results for the fusion of
two identical spherical aggregates. We show that in both cases the computed
$(r/R_0)^2$ vs $t/\tau$ dependence can be reasonably well fitted by
Eqs.~(\ref{exp_fit_formula}). Then, using experimental results for aggregate
fusion \cite{jakab08-2438}, the calibration of the KMC and CPD simulation time
scales is exemplified for the case of cardiac cushion tissue (CT). Finally,
KMC and CPD simulations are used to predict the formation of a toroidal
structure by cell aggregate fusion, an important structure in the engineering
of tubular tissue constructs \cite{jakab08-413}.

\subsubsection{KMC simulations}
\label{sec:kmc-fusion}
The initial radius of the two identical fusing aggregates used in our KMC
simulation was $R_0 = 10$ cell diameters. Each aggregate contained 5,927
cells, with a cell-cell work of cohesion $ \epsilon_{\rm cc} = 0.9$. The
medium-medium (cell-medium) work of cohesion (adhesion),
$\epsilon_{\text{mm}}$ ($\epsilon_{\text{cm}}$), was considered to be
negligibly small.
A total of 10 KMC simulations of the same fusion process were carried out, each
time using a different seed of the random number generator. Each simulation
was run for $10^5$ KMC time steps.

\begin{figure}
\includegraphics[width=3.2in]{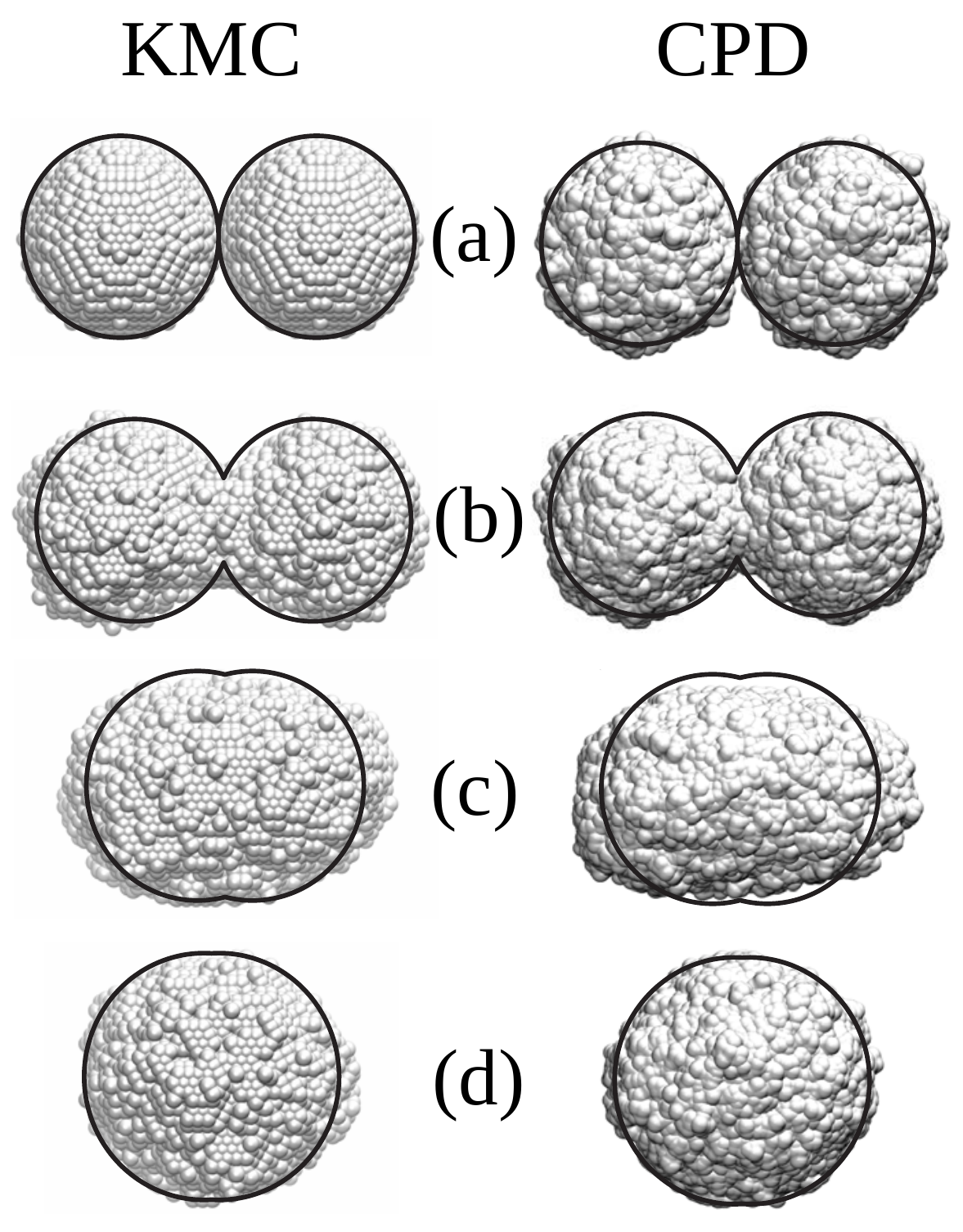}
\caption{\label{fusion_snap_kmc} Time evolution of the fusing aggregates in
  the KMC (left) and CPD (right) simulations. The snapshots were taken at: (a)
  $t = 0$, (b) $t = 0.19 \tau$, (c) $t = 2.8 \tau$, and (d) $t = 5.5
  \tau$. The solid-line contours represent the theoretical shapes of the fusing
  aggregates determined by Eqs.~(\ref{exp_fit_formula}). }
\end{figure}

\begin{figure}
\includegraphics[width=3.25in]{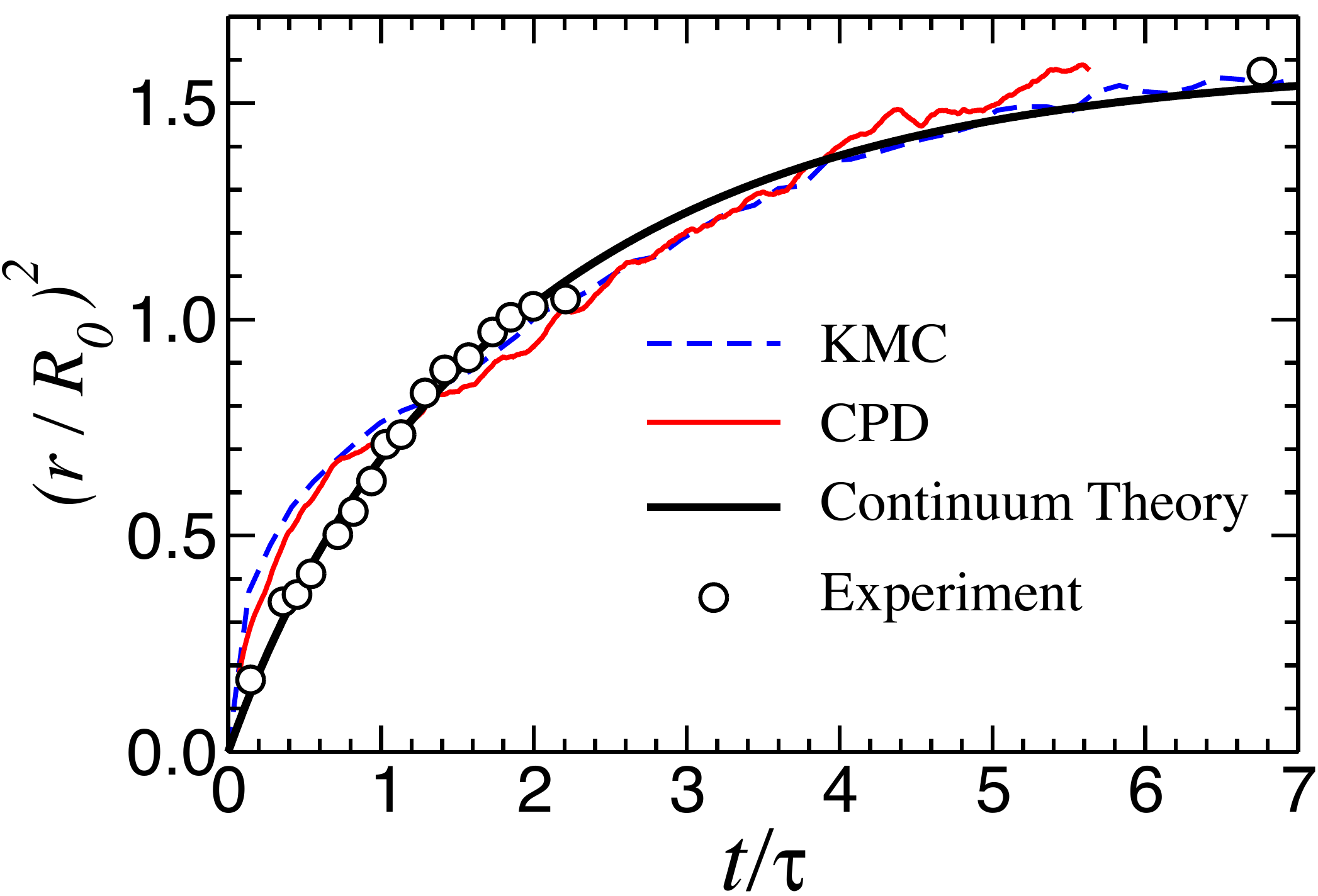}
\caption{\label{cross_area_all}(color online). Comparison of $(r/R_0)^2$ vs
  $t/\tau$ for the fusion of two spherical aggregates obtained from KMC
  simulations (dashed line), CPD simulations (thin solid line), continuum
  theory (thick solid line) and experiment (circles) using cardiac cushion
  tissue (CT) aggregates \cite{jakab08-2438}.}
\end{figure}

Representative snapshots during the KMC fusion simulation are shown in
Fig.~\ref{fusion_snap_kmc}. 
The corresponding $(r/R_0)^2$ vs $t/\tau$
dependence is shown in Fig.~\ref{cross_area_all} (dashed curve). Apart from
the beginning of the fusion process (i.e., $t<\tau$) the KMC result appears to
match rather well both the theoretical prediction (thick-solid curve),
Eqs.~(\ref{exp_fit_formula}), and the experimental results corresponding to the
fusion of CT aggregates (open-circle) \cite{jakab08-2438}. 

In the KMC time unit $t_0=w_0^{-1}$, the fusion time [obtained by fitting the
KMC simulation results to the theoretical formula
Eqs.~(\ref{exp_fit_formula})] was $\tau_0=1.1\times 10^9$.  Since the
experimental characteristic fusion time for CT aggregates
$\tau_{\text{exp}}\approx 5$h \cite{jakab08-2438}, it follows that the KMC
time unit (for CT aggregates used in \cite{jakab08-2438}) has the calibrated
value $t_0=w_0^{-1} = \tau_{\text{exp}}/\tau_0=1.6\times 10^{-5}$s.

\subsubsection{CPD simulations}
\label{sec:cpd-fusion}
Each of the two spherical aggregates used in the CPD simulation of aggregate fusion contained $2000$ cells.  
{\color{black} The used CPD parameters and integration timestep, in CPD units, are described in Sec.~\ref{sec:cpd-units}.}
The equilibrated aggregates were placed within a distance of one $\sigma$ before starting the fusion simulation.

Representative snapshots during the fusion process are shown, and compared
with the corresponding KMC simulation results, in
Fig.~\ref{fusion_snap_kmc}. While in both KMC and CPD simulations the profiles
of the fusing aggregates for intermediate stages of the fusion process
(Fig.~\ref{fusion_snap_kmc}b-c) agree quite well, these show noticeable
differences with respect to the theoretical prediction,
Eqs.~(\ref{exp_fit_formula}), shown as solid-line contours in
Fig.~\ref{fusion_snap_kmc}.

The $(r/R_0)^2$ vs $t/\tau$ dependence in the CPD simulation is also shown in
Fig.~\ref{cross_area_all}. The CPD and KMC simulation results are similar. Apart from short times ($t<\tau$) they agree quite well with both the theoretical prediction, Eqs.~(\ref{exp_fit_formula}), and the experimental results for CT \cite{jakab08-2438}.

The characteristinc CPD fusion time is determined to be $\tau\approx 540 t_0$. By equating this with $\tau_{\text{exp}}\approx 5$h, one finds that the CPD time unit calibrated for CT aggregates is $t_0\approx 0.6$~min. 
{\color{black} The CPD simulations were preformed on 32 CPUs of a dual core 2.8GHz Intel Xeon EM64T cluster with a performance of around 5 million timesteps/day (which is equivalent to $500 t_0$ and slightly less than $1 \tau$).}

\subsubsection{Toroidal structure formation}
\label{sec:toro}

\begin{figure}
\includegraphics[width=3.2in]{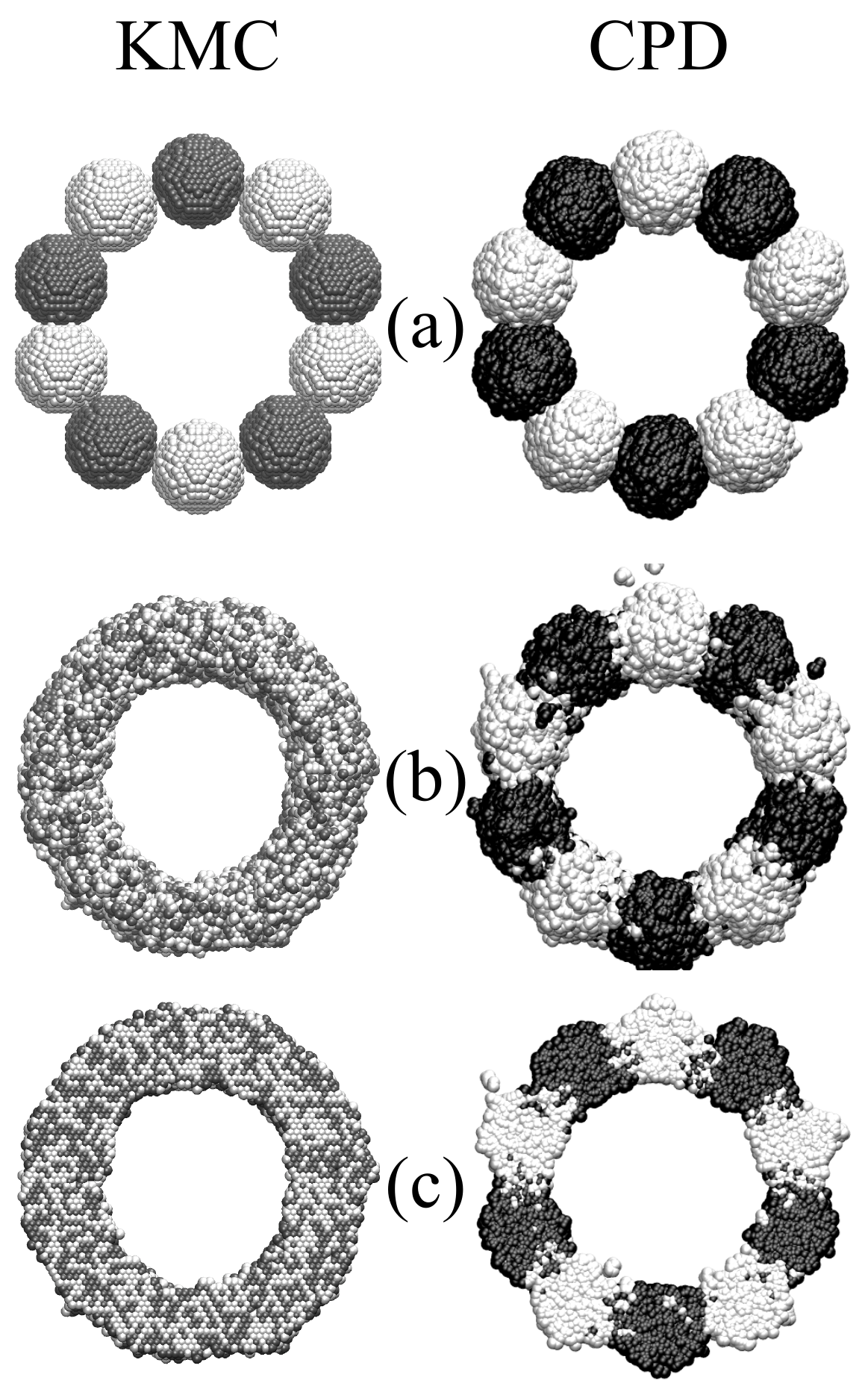}
\caption{\label{snap_lifesaver} KMC (left) and CPD (right) simulations of
  toroidal structure formation through the fusion of $10$ cell aggregates. Top view
  of the fusing aggregates at (a) the beginning ($t=0$), and (b) the
  completion of fusion. (c) Cross-section through the median plane of the
  fused toroidal structure shown in (b). Otherwise identical cells, initially
  located in adjacent aggregates are colored differently to emphasize the
  degree of mixing during fusion. }
\end{figure}

Once the KMC and CPD time scales have been calibrated from the fusion of two spherical CT aggregates, one can employ KMC and CPD simulations to describe and predict the time evolution of more complex CT structures, which are not tractable analytically. To exemplify this point, here we consider the formation of a toroidal structure as a result of the fusion of $10$ identical CT spherical aggregates initially arranged in a circular configuration as shown in Fig.~\ref{snap_lifesaver}a. The corresponding KMC and CPD simulations were carried out using the same model parameters as in the fusion of two aggregates described above. In both KMC and CPD simulation the fusion process into a toroidal ring appeared to be completed in $\Delta{t}\approx 2.5\tau\approx 12.5$~h, as shown in Fig.~\ref{snap_lifesaver}b.
{\color{black}This prediction should be easily testable experimentally.}

While it seems that both KMC and CPD methods are capable of providing a fairly
good description of the shape evolution of a multicellular system during its
biomechanical relaxation process, the actual cellular dynamics in the two
methods is quite different. Indeed, unlike in CPD simulations, in KMC
simulations the motion of individual cells is unrealistically fast. This point
is manifest in Fig.~\ref{snap_lifesaver}. By the time the toroidal ring
structure is formed, in the KMC simulation, cells from adjacent aggregates
(colored differently) appear to be completely mixed. This is clearly not the
case in the CPD simulations, where, similarly to existing experimental results
\cite{jakab08-2438,jakab08-413}, there is little mixing between the cells
of the fused adjacent aggregates. 

\begin{figure}
\includegraphics[width=3.25in]{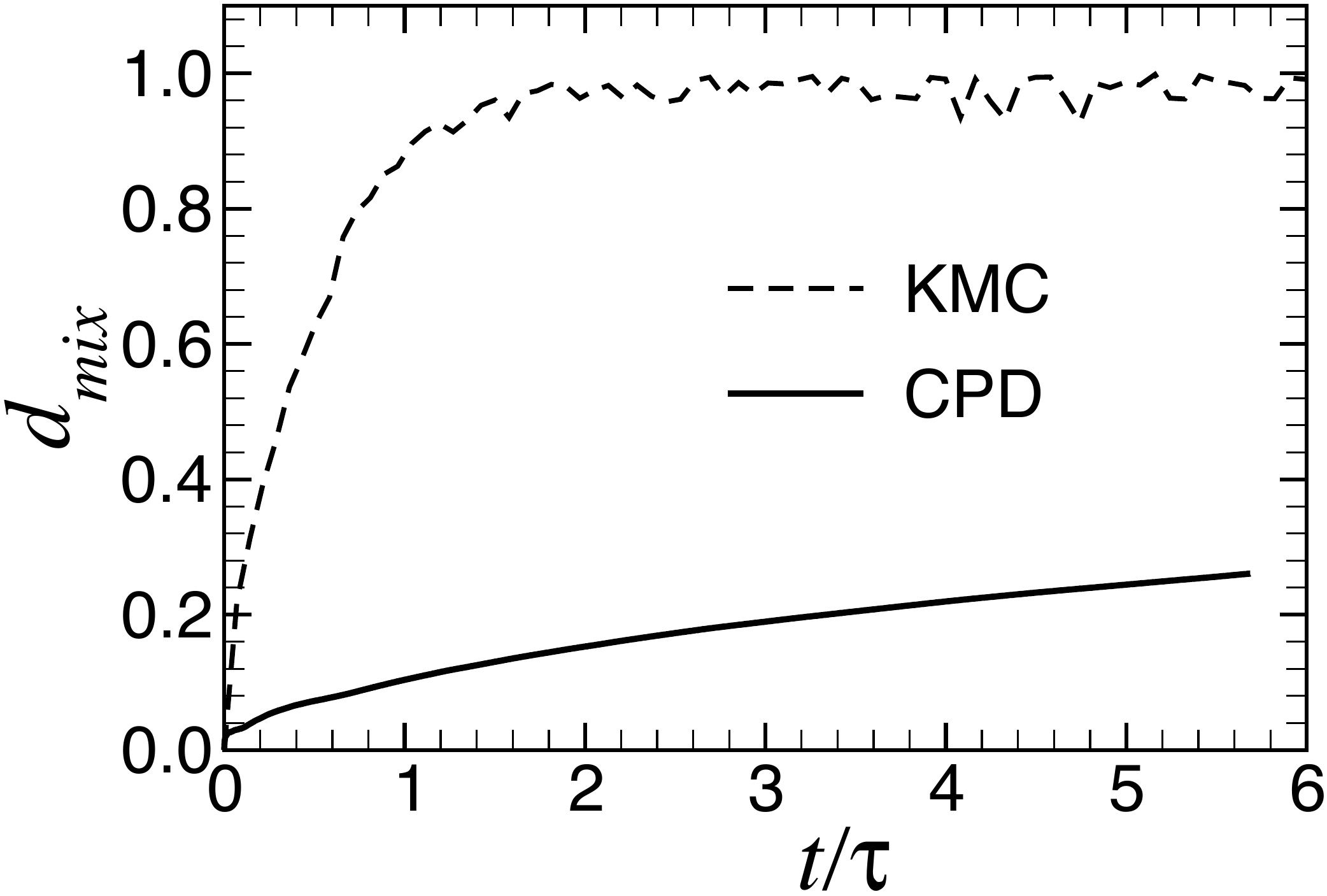}
\caption{\label{mixing-param-cpd} Time evolution of the mixing parameter
  $d_{mix}$ calculated for the fusion of two cellular aggregates from the CPD
  (solid line) and KMC (dashed line) simulations. } 
\end{figure}

To further emphasize this point, we have quantified the degree of cellular
mixing during the fusion, along the $x$-axis, of two identical spherical
aggregates [labeled as $L$ (left) and $R$ (right)], with initial radius $R_0$
(see Fig.~\ref{caps}), by calculating the time dependent mixing parameter
\begin{equation}
  \label{eq:dmix}
  d_{mix}(t) = \frac{4}{M}\sum_{m=1}^M \frac{\Delta{N_m^L(t)}\cdot
    \Delta{N_m^R(t)}}{[\Delta{N}_m(t)]^2} .
\end{equation}
Here $\Delta{N_m^L}(t)$ [$\Delta_m^R(t)$] is the number of CPs situated
initially (at $t=0$) in the $L$ ($R$) aggregate and having, at time $t$, the
$x(t)$ coordinate in the interval $\{-2R_0+(m-1)\Delta{x}, -2R_0+m\Delta{x}\}$,
$1 \le m \le M$, with $M$ a properly chosen, sufficiently large integer,
$\Delta{x}=4R_0/M$, and $\Delta{N_m(t)}=\Delta{N_m^L}(t)+\Delta{N_m^R}(t)$.
Clearly, $d_{mix}$ can take values between $0$ (completely unmixed system)
and $1$ (uniformly mixed system).  

The time evolution of $d_{mix}(t)$ is shown in Fig.~\ref{mixing-param-cpd}.  In
the KMC simulation cellular mixing is almost complete ($d_{mix}=1$) after the
characteristic fusion time $\tau$, i.e., significantly sooner than the
completion of the fusion process ($\sim 6\tau$). By contrast, in the CPD
simulation even at the end of the fusion $d_{mix}\sim 0.2\ll 1$. Based on
these results one may conclude that: (i) the cellular dynamics that drives
aggregate fusion in the KMC simulations is unrealistic (i.e., the system is
too liquid-like), and (ii) the CPD model provides a more realistic and
atractive approach to describe biomechanical relaxation processes of
multicellular systems.

\subsection{Cell Sorting in Two Component Aggregates}
\label{sec:sorting}
When two populations of cells of different adhesivities are randomly mixed 
within a multicellular aggregate, they sort such that the more adhesive cells
occupy the internal region while being surrounded by the less adhesive
cells. Cell sorting has been extensively studied both \textit{in vitro}
\cite{steinberg62-1577,steinberg63-257,steinberg70-395,perez-pomares06-809}
and \textit{in silico} \cite{graner92-2013,glazier93-2128, mombach95-2244}.

According to DAH, the outcome of cell sorting in a two-component multicellular
aggregate (composed of two types of cells, labeled `$a$' and `$b$') depends on
the relative magnitude of the corresponding works of cohesion/adhesion needed
to separate cells of the same/different types (i.e., $\epsilon_{aa}$,
$\epsilon_{bb}$, and $\epsilon_{ab}$), respectively \cite{foty04-397}.
Here we employ both KMC and CPD simulations (described in Secs.~\ref{sec:kmc}
and \ref{sec:cpd}) to investigate cell sorting in a spherical aggregate of two
cell types $a$ and $b$, with $\epsilon_{aa}<\epsilon_{bb}$. We consider three
cases, referred to as C1, C2 and C3, that lead to qualitatively different
experimental outcomes \cite{foty04-397}. C1: For intermediate adhesion
between $a$ and $b$ cells, i.e., $\epsilon_{aa} < \epsilon_{ab}<
(\epsilon_{aa}+\epsilon_{bb})/2$, the less cohesive $a$ cells engulf the more
cohesive $b$ cells, thus leading to the complete segregation (see Fig.~\ref{snap_sort_kmc}b). C2: For strong $a$--$b$ adhesion,
i.e., $(\epsilon_{aa}+\epsilon_{bb})/2< \epsilon_{ab}$, there is limited
sorting and the spherical aggregate remains more or less homogeneously mixed
(see Fig.~\ref{snap_sort_kmc}c). C3: For weak $a$--$b$ adhesion, i.e.,
$\epsilon_{ab}<\epsilon_{aa} <\epsilon_{bb}$, the two types of cells
completely separate by transforming the initial spherical aggregate into two
attached homogenous spheroidal caps (each containing either $a$ or $b$ cells)
as shown in Fig.~\ref{snap_sort_kmc}d. Thus, the degree of
cell sorting is enhanced (reduced) for small (large) values of the adhesion
energy $\epsilon_{ab}$, compared to the corresponding cohesion energies
$\epsilon_{aa}$ and $\epsilon_{bb}$.
Note that in terms of the interfacial tension $\gamma_{ab}$ (defined below
Eq.~(\ref{eq:totalenergy}), for $``1"=a$ and $``2"=b$), case C1
corresponds to $\gamma_{ab}>0$ and $\epsilon_{ab}>\epsilon_{aa}$, while case
C2 corresponds to $\gamma_{ab}<0$. The inequalities defining case C3 also
imply $\gamma_{ab}>0$. Thus, in a multicellular aggregate with two types of
cells, in order to have cell sorting (segregation) the corresponding
interfacial tension must be positive (i.e., $\gamma_{ab}>0$). The larger this
parameter the more efficient and complete the sorting.

The results of our KMC and CPD simulations, presented next, appear to be in
good agreement with \textit{in vitro} experimental findings for these three
cases \cite{foty04-397}.

\subsubsection{KMC simulations}
\label{sec:kmc-sorting}
\begin{figure}
\includegraphics[width=3.2in]{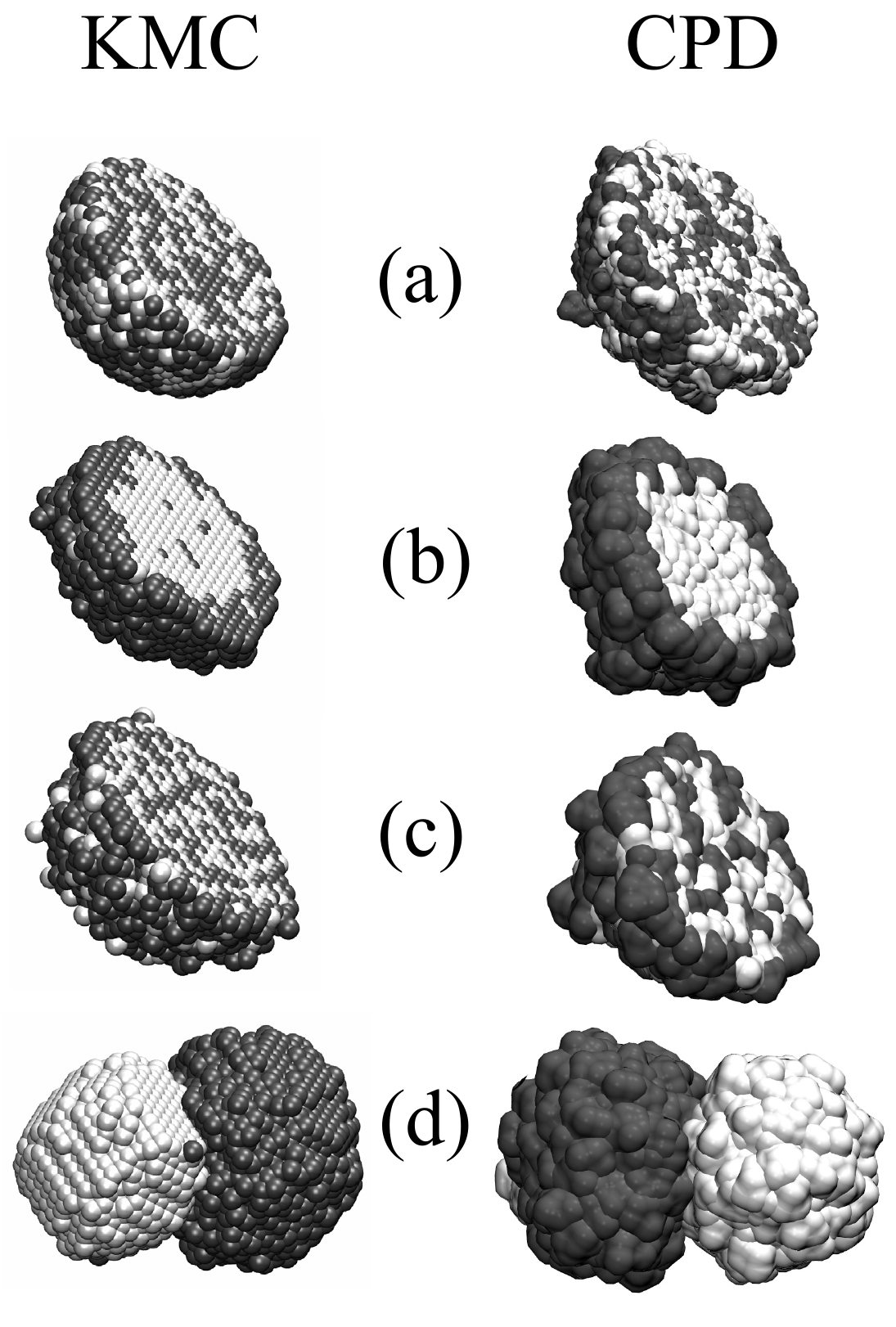}
\caption{\label{snap_sort_kmc} 3D snapshots from KMC (left) and CPD (right)
  simulations of the cell sorting in an initially spherical aggregate
  composed of two randomly mixed cell types (black and light
  grey). The snapshots represent the (a) initial, and (b-d) final
  configurations of the simulated system. The latter correspond to (b)
  intermediate (case C1), (c) strong (case C2), and (d) weak (case C3) cell
  adhesion energy, as explained in the text. For better
  visualization of cell mixing/sorting in (a)-(c) only half of the spherical
  aggregate is shown. (Images rendered with VMD \cite{humphrey96-33}).}
\end{figure}

We have performed three KMC simulations of cell sorting starting with a
spherical aggregate composed of a random mixture of $N_a=3,589$ less cohesive
cells of type $a$ and $N_b=2,362$ more cohesive cells of type $b$ (i.e., with
$\epsilon_{aa}<\epsilon_{bb}$). Thus, the spherical aggregate had a total of
$N=5,951$ cells, and a radius of about $10$ cell diameters.
The values of the model parameters used in the three KMC simulations,
corresponding to cases C1, C2 and C3 described above, are listed in
Table~\ref{paramtable}. Each KMC simulation was performed up to $10^5$
(non-uniform) time steps, given by Eq.~(\ref{eq:dt}), leading to the 
final configurations shown in
Fig.~\ref{snap_sort_kmc}b-d.

\begin{table}
  \caption{\label{paramtable} Values of the model parameters (energies
    expressed in units of $E_T$) used in the KMC and CPD simulations shown in Fig.~\ref{snap_sort_kmc}.} 
\begin{ruledtabular}
\begin{tabular}{ccccrcc}
Simulation & $\epsilon_{aa}$ & $\epsilon_{ab}$ & $\frac{\epsilon_{aa} +
\epsilon_{bb}}{2}$ & $\gamma_{ab}$ & Case & Outcome\\ 
\hline
KMC & 1.0 & 1.1 & 1.4 &  0.3 & C1 & Fig.~\ref{snap_sort_kmc}b left\\
KMC & 1.0 & 1.5 & 1.4 & -0.1 & C2 & Fig.~\ref{snap_sort_kmc}c left\\
KMC & 1.0 & 0.3 & 1.4 &  1.1 & C3 & Fig.~\ref{snap_sort_kmc}d left\\
CPD & 0.8 & 0.9 & 1.0 &  0.2 & C1 & Fig.~\ref{snap_sort_kmc}b right\\
CPD & 0.8 & 1.1 & 1.0 & -0.1 & C2 & Fig.~\ref{snap_sort_kmc}c right \\
CPD & 0.8 & 0.2 & 1.0 &  0.8 & C3 & Fig.~\ref{snap_sort_kmc}d right \\
\end{tabular}
\end{ruledtabular}
\end{table}

To quantify the degree of cell sorting as a function of
time during the KMC simulations, we used a sorting parameter $s$ defined
as  \cite{palsson01-835}
\begin{equation}
  \label{eq:s}
  s = \frac{1}{N}\sum_{i=1}^{N}\frac{N_{t_i}}{N_i} ,
\end{equation}
where $N$ is the total number of cells in the system, and for a given cell
$i$, $N_i$ ($N_{t_i}$) is the number of nearest neighbor cells regardless
of their type (of the same type $t_i$ as the cell $i$). The sum in
Eq.~(\ref{eq:s}) runs over all cells in the system.  
Clearly, $0<s<1$, and the larger $s$ the more complete the sorting. Note
that even for completely sorted multicellular systems, built from two (or
more) different cell types, the presence of the interface(s) between the
segregated regions renders the maximum possible value, $s_{max}$, of the
sorting parameter $s_{max}<1$. 
For example, in the above case C1, when at the end of sorting $N_a$ cells of
type $a$ completely engulf $N_b$ cells of type $b$, one can estimate $s_{max}$
as follows. For simplicity, assume that both cell types have spherical shape
with the same diameter $d$. Let $\Delta{N}$ be the number of cells (of either
type $a$ or $b$) situated at the spherical interface, of mean radius $R_b$ and
width $\Delta{R}$, between the two segregated regions (see
Fig.~\ref{snap_sort_kmc}b), and $N=N_a+N_b$. Since for a cell $i$ situated at
the interface $N_{t_i}/N_i\approx 1/2$, according to Eq.~\eqref{eq:s},
\begin{equation*}
s_{max} \approx \frac{1}{N}\left[\frac{1}{2}\times\Delta{N} + 1\times
  (N-\Delta{N}) \right]   
  = 1-\frac{1}{2}\frac{\Delta{N}}{N}  .
\end{equation*}
Furthermore, assuming that cells are distributed uniformly within the
aggregate, one has $N_b (d/2)^3\approx R_b^3$, i.e., $R_b\approx N_{b}^{1/3}
d/2$, and $\Delta{N}\times (4\pi/3) (d/2)^3 \approx 4\pi R_b^2\Delta{R}$,
implying $\Delta{N} \approx 6 N_b^{2/3} (\Delta{R}/d)$. Finally, assuming that
the thickness of the interfacial layer, separating the segregated cell
regions, is $\Delta{R} = x d$, where $2<x<3$, one obtains
\begin{equation}
  \label{eq:smax}
  s_{max} \approx 1-3x\frac{N_b^{2/3}}{N} . 
\end{equation}
Note that according to Eq.~\eqref{eq:smax}, as $N\rightarrow\infty$, i.e., for
large aggregates, $s_{max}$ approaches unity as $N^{-1/3}$ (assuming that
$N_a$ and $N_b$ are of the same order of magnitude).

\begin{figure}
\includegraphics[width=3.2in]{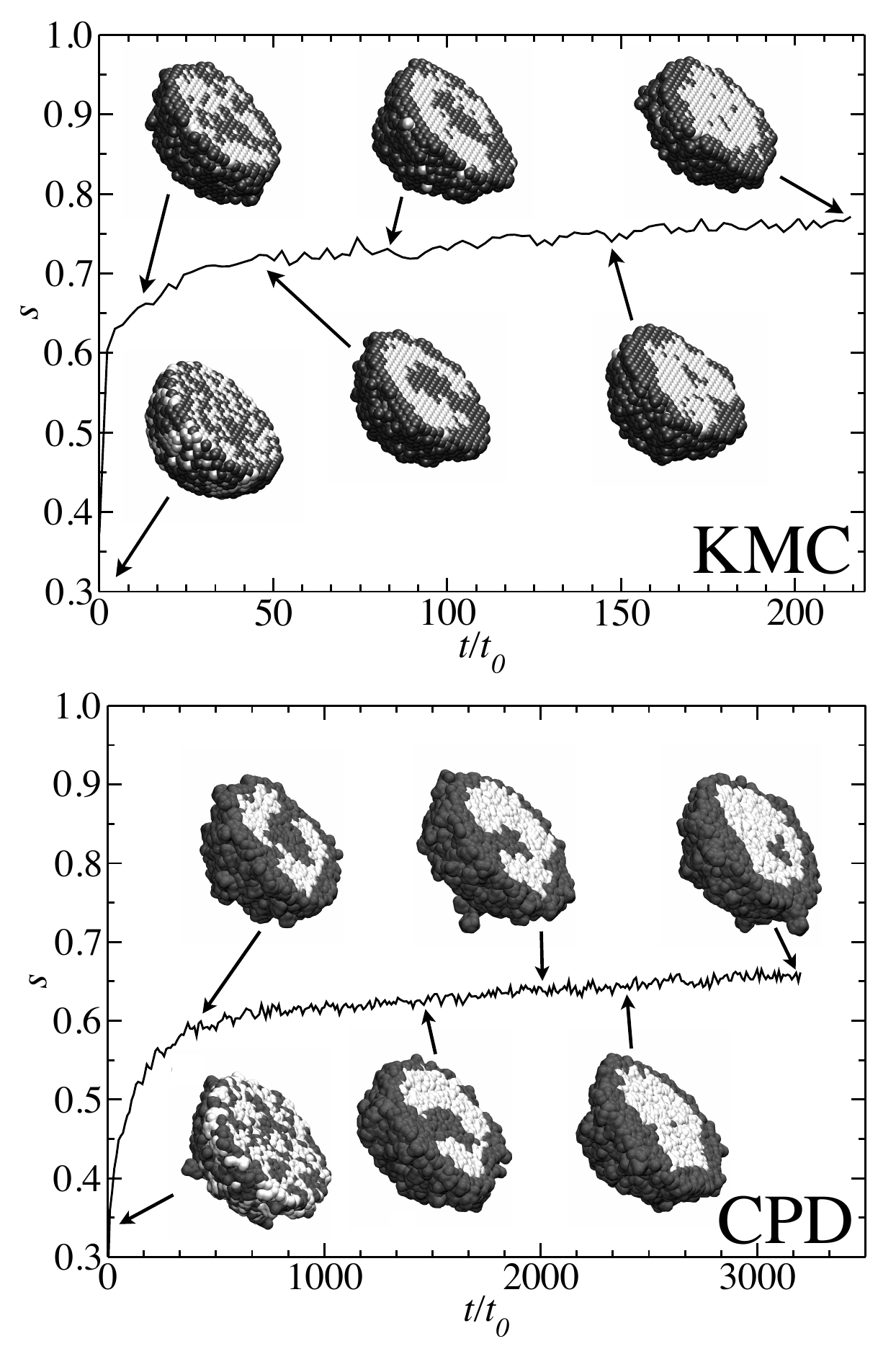}
\caption{\label{snap_sort_param_kmc}   
  Time dependence of the cell sorting parameter, $s=s(t)$, corresponding to
  case C1 described in the text, for both KMC (top) and CPD (bottom)
  simulations. The insets represent snapshots of half of the spherical
  aggregate taken at times indicated by arrows.  }
\end{figure}

The time evolution of the sorting parameter, $s=s(t)$, in our KMC simulation
corresponding to case C1 is shown in Fig.~\ref{snap_sort_param_kmc}. The insets represent snapshots of the sorting process taken at times indicated by the arrows. 

The sharp increase of $s(t)$ at the beginning of the simulation followed by a
slow asymptotic approach to $s_{max}$ indicates that there are at least two
sorting time scales. Indeed, the entire time evolution of the sorting
parameter can be well fitted with the double exponential
\begin{equation}
  \label{eq:s-fit}
  s(t) = s_{max} - s_1e^{-t/\tau_1} - s_2e^{-t/\tau_2},
\end{equation}
where $s_{max}=0.76$ is in very good agreement with the theoretically
estimated value $0.78$ obtained from Eq.~\eqref{eq:smax} for $x=2.5$. The
other fitting parameters in Eq.~\eqref{eq:s-fit} are: $\tau_1=1.4 t_0$,
$s_1=0.27$, $\tau_2=58.5 t_0$ and $s_2=0.11$. The shorter time scale $\tau_1$
corresponds to the local rearrangement (sorting) of cells leading to small
clusters of same types of cells, while the longer time scale $\tau_2$ describes
the much slower engulfment process of the $b$ cells by the $a$ cells, a
process that requires large displacements by a finite number of cells.

Although the results of our KMC simulations appear to be in good qualitative
agreement with experiments on cell sorting
\cite{steinberg63-401,steinberg70-395,foty04-397}, a quantitative comparison,
e.g., in terms of the time evolution of the sorting parameter, is not feasible
because $s(t)$ cannot be measured experimentally. Thus, there is no simple way
to reliably calibrate the time unit $t_0$ (which is related to the model
parameter $w_0$) used in the plot of $s$ \emph{vs} $t/t_0$ in
Fig.~\ref{snap_sort_param_kmc}.
However, $s(t)$ can also be determined from CPD simulations, thus allowing for
a quantitative comparisson between the two computer simulation methods.

\subsubsection{CPD simulations}
\label{sec:cpd-sorting}

We have also used CPD simulations to investigate cell sorting corresponding
to the three cases C1, C2 and C3 described above.
The initially spherical aggregate contained a random mixture of equal number
$N_a=N_b=1,000$ of cells of type $a$ and $b$. 
While the CPD parameters $\epsilon_{aa} \equiv \epsilon_a^{inter} =
\epsilon_a^{intra} = 0.8$ and $\epsilon_{bb} \equiv \epsilon_b^{inter} =
\epsilon_b^{intra} = 1.2$ were kept the same in all three simulations, the
parameter $\epsilon_{ab} \equiv \epsilon_{ab}^{inter}$ had different values
(similar to the ones used in the KMC simulations) for the three cases C1, C2
and C3 as listed in Table~\ref{paramtable}.
The cell sorting patterns obtained at the end of the corresponding CPD
simulations are shown in Fig.~\ref{snap_sort_kmc}. As expected, these patterns
are similar to the ones obtained in the KMC simulations.

In order to quantify the degree of cell sorting in the CPD simulations by
employing the cell sorting parameter $s$, defined through Eq.~\eqref{eq:s}, we
determined the position of a cell by the center of mass of the constituent
CPs, and considered two cells to be neighbors if they were separated by a
distance less than 3.25~$\sigma$. For the CPD simulation corresponding to case
C1, $s(t)$ is shown Fig.~\ref{snap_sort_param_kmc}. Similarly to the KMC result,
$s(t)$ can be fitted well with the double exponential \eqref{eq:s-fit}. Again,
$s_{max}=0.68$ is in good agreement with the theoretical prediction
Eq.~\eqref{eq:smax}, i.e., $0.67$ for $x=2.2$ (or $0.63$ for $x=2.5$). The
other fitting parameters in Eq.~\eqref{eq:s-fit} are: $\tau_1=0.68~t_0$,
$\tau_2=103~t_0$, $s_1=0.25$ and $s_2=0.1$. Note that while $s_1$ and $s_2$
have essentially the same values for both KMC and CPD simulations, the time
constants $\tau_1$ and $\tau_2$ are quite different, as the
corresponding time units $t_0$ are different in the two simulations.
Moreover, the fact that, for similar model parameters, $\tau_2/\tau_1=41.8$ in
KMC is about twice as large as $\tau_2/\tau_1=21.7$ in the corresponding CPD
simulation indicates that the self diffusive motion of cells in KMC occurs
much faster than in CPD. In other words, the multicellular system is more
liquid-like in KMC than in CPD simulations.

\section{Conclusions}
\label{sec:conclusions}
We have presented two (KMC and CPD) simulation methods and an analytic, continuum theoretical approach to address structure formation by cell sorting and the fusion of contiguous multicellular spheroids. The theoretical method was used to interpret the experimental results, and to test the viability and calibrate the model parameters of the KMC and CPD simulations.
Our study was motivated by the need to quantify biomechanical properties of engineered tissue constructs, composed of compact tissues made of adhesive and motile cells and to predict their time evolution. The growing interest for understanding shape changes in such tissue constructs stems from their applications in tissue engineering in general and in the emergent field of 3D bioprinting in particular \cite{jakab08-413}.

The KMC method is based on a lattice representation of the 3D tissue construct
and dynamics is described in terms of rates associated with possible movements
of cells.  Similarly to previously employed MMC studies, the mixing pattern
observed in KMC simulations disagrees with experiments. In both methods an
elementary move consists in cells swapping positions with neighbors, which
overestimates cell motility. Nevertheless, the time scale calibration in KMC
makes the simulated time course realistic as far as the shape evolution of
multicellular tissue constructs is concerned.

The CPD method is based on modeling individual cells in a tissue construct as
interacting CPs. The dynamics of the multicellular system are determined by
integrating the equations of motion for each CP.
{\color{black}The CPD force field parameters are determined such that the time evolution of the shape of the fusing spherical aggregates in the CPD simulation matches as closely as possible the experimental one (i.e., two touching spherical caps). Once the CPD model is calibrated, this can be used to simulate the shape evolution of arbitrary 3D multicellular constructs. 
It should be emphasized that in CPD (i.e., computer) units the calibrated CPD parameters (and therefore the outcome of a CPD simulation) are independent of the used cell type. However, the CPD units \eqref{eq:cdp-units} have specific values for different cell types. Thus, the CPD simulations reported here can be applied as is to different cell types; the corresponding CPD time unit $t_0$ should be determined in each case by equating $\tau_{sim}\approx540\, t_0$ (see Sec.~\ref{sec:cpd-fusion}) with the experimental fusion time $\tau_{exp}$. }
The reported CPD simulations provided a good description for both fusion and cell sorting of multicellular spheroids. We found that CPD provides a more realistic description of complex multicellular structure formation than KMC. Indeed, the behavior of the studied multicellular systems in CPD simulations resembles to that of complex visco-elastic materials while in KMC simulations to that of viscous liquids. It is to be expected that by including more realistic features into the interaction of the CPs the accuracy of the CPD method can be further improved.

\begin{acknowledgments}
  This work was supported by grants from the National Science Foundation
  (PHY-0957914 and FIBR-0526854). The work by A.N. was supported in part by
  the Romanian National Authority for Scientific Research (CNCSIS Contract
  PCCE-ID 76).  Computational resources were generously provided by the
  University of Missouri Bioinformatics Consortium.
\end{acknowledgments}


%

\end{document}